\documentclass{article}       

\usepackage{graphicx}
\usepackage[english, ngerman]{babel}
\usepackage[T1]{fontenc} 
\usepackage{amsfonts}
\usepackage{amssymb}
\usepackage{amsmath}
\usepackage{graphicx}
\usepackage{multicol}  
\usepackage{hyperref} 
\usepackage{eurosym}

\begin{document}

\title{\flushleft\normalsize GARBE, ANNETTE. \textit{Die partiell konventional, partiell empirisch bestimmte Realit"at physikalischer RaumZeiten.} K"onigshausen \& Neumann: W"urzburg 2001, pp. 419, \euro\, 51,00, ISBN: 978-3-8260-2080-3}

\date{}

\maketitle  

\noindent
In 2001 the book \textit{Die partiell konventional, partiell empirisch bestimmte Realit"at physikalischer RaumZeiten} was published by K"onigshausen \& Neumann, with a volume of more than 400 pages. It was written by Annette Garbe.\footnote{It was Annette Garbe's "`Doktorarbeit"' (in English: Ph.D.thesis). The doctoral supervisor and first advisor was Wolfgang H. Schrader (Department of Philosophy, University of Siegen), the second advisor was Claus Grupen (Department of Physics, University of Siegen). Professor Schrader died in 2000, shortly after Garbe had completed her doctoral thesis and had past the oral examination. Professor (Emeritus) Grupen told me he can still remember well that Professor Schrader felt overburdened to write an assessment over the work due to the discussed mathematical and physical topic that requires at least some familiarity with analysis and differential equations. So, it was Grupen who wrote the assessment. To my question as to why the book is not quoted, but wholly ignored by the scientific community, Grupen answered: ``I suspect that the text is just too difficult to read for philosophers.'' One might add: ``And too philosophical for physicists.'' I gratefully acknowledge Claus Grupen for his kindness answering my questions.} The English translation might be: \textit{The Partial Conventionally, Partial Empirically Determined Reality of Physical Space-Times}.

In the introduction Garbe gives a very telling historical overview of non-Euclidean geometry. It focuses on the impossibility proof of the parallel postulate on the basis of the other postulates of Euclidean geometry and emphasizes the importance of relative consistency proofs. 

The following chapter of Garbe's book deals with the problem of the foundations of geometry. It begins with Karl Friedrich Gauss' ``test triangles,'' i.\,e. the measure of the angles of the great triangle formed by the mountain peaks in Germany, Hohenhagen, near G"ottingen, Brocken, in the Harz Mountains, and Inselberg, in the Th"uringer Wald to the south. If Gauss had been able to take measurements of sufficient accuracy, he might have found that the sum of the three angles of his great triangle differed from two right angles by an amount. There is no clear documentary evidence that Gauss was actually seeking evidence of non-Euclidean geometry of physical space. As Garbe points out, the measurements of the angles of the triangle formed by the three peaks of Hohenhagen, Inselberg and Brocken, which Gauss made during his geodetic survey of the Kingdom of Hannover, were not intended as an experimental test of the Euclidean nature of the geometry of physical space.  We now know that there is an angular discrepancy due to relativistic effects, but it is far too small to be directly observed. So, the question if it is possible to decide empirically whether physical space is Euclidean is not simply been plucked out of the air.

At this point, Garbe compares Henri Poincar\'e's conventionalism with David Hilbert's formalism in detail, taking into account Bernhard Riemann's, Hermann von Helmholtz' and Moriz Pasch's works. The crucial point on which Garbe refers to is the often discussed problem of the relation between logic, mathematics and physics, and the problem of the possibility of deciding which axiom system is appropriate to the representation of physical space and time. 

As well known, Hans Reichenbach argued that a geometry can be true or false only relative to the coordinative definitions which have been laid down beforehand as conventions. That is, the statements of geometry are empirical provided that the coordinative definitions for geometry are laid down as conventions. These coordinative definitions are to determine how geometrical quantities, such as length, are to be measured and how geometrical relations, like those of congruence, are to be ascertained.
 
Garbe picks up on this idea and generalizes it: One cannot determine the metric structure of space-time without knowing the physical laws underlying them. Vice versa, one cannot formulate the physical laws without knowing the underlying space-time structure. We are facing a vicious circle. 

Garbe maintains that this circle can be broken by way of conventional chosen definitions. They act, so to say, a priori in a very flexible manner. One should not confuse conventions with arbitrary assertions. Experience plays a crucial role in determining and evaluating conventions. This is the basic thesis that Garbe tries to explain and defend in her book, called ``partial determination.''

It is not a coincidence that Annette Garbe borrows the term ``partially determined'' from Ludwig Lange, who is known for the introduction of the concept of inertial reference frames in 1885. Lange explored the question how one has to proceed in order to avoid circular reasoning regarding to the basic concepts and laws of physics. Generally speaking, a \textit{petitio principii} occurs when one attempts to infer a conclusion that is based upon a premise that ultimately contains the conclusion itself, i.\,e. the proposition to be proved is assumed (implicitly or explicitly) in one of the premises. Or, a term is definied by using the term in the definition. 

Garbe discusses a number of examples for circular reasoning in physics which are listed by Lange, e.\,g. the law of the relatively invariable inclination of all gyroscopic rotation axes had to be stated in advance of the definition of the fundamental system, and therefore also of the law of inertia. One could not do other than to derive (implicitly) this theorem, now in the converse direction, from the law of inertia. However, this would be a circular derivation, comparable to the one in geometry, eliminated long ago, as Lange added, which contains the unreasonable demand to define the straight line as the shortest path between two points, and to prove afterwards that the straight line must be the shortest path between two points. Especially, Lange criticized the ``scientific dissonance'' with respect to the definition of equal time intervals. In a first step, equal time intervals were defined by establishing that identical processes take place during these intervals. In a second step, we observe then that identical processes take equal time intervals. 

However, as Garbe emphasizes, Lange offers a way out. Lange called it ``principle of particular determination'' (German: "`Princip der Particulardetermination"'). It rests on the idea that every formulation of a physical law includes a conventional part. Thus, in the expression of the principle of inertia, we can rely at first merely on a conceptual system of reference, to which we describe all the determinations as required. By means of conceptual definitions we stipulate a spatial inertial system and an inertial timescale, and take both as the basis for all further considerations on the phenomena and their reciprocal relations. 

Lange's proposal has to bee seen in the context of the long-standing and lively debate on absolute space, time and motion. Fitting hand in glove with the contemporary work of others (e.\,g. Carl Neumann, Heinrich Streintz, James Thomson, Ernst Mach, Ludwig Boltzmann, Heinz Kleinpeter, Paul Volkmann, Gottlob Frege et al.) at the end of the 19th century, Lange pleaded for the elimination of absolute concepts from physics.  The original question, relative to what frame of reference the laws of motion hold is revealed to be wrongly posed. Physical laws essentially determine a class of reference frames, and (in principle) a procedure for constructing them. The idea is to consider the laws as regularities, or ``rules of research'' (German: ``Forschungsregeln'') that systematize and describe the space-time structure. Consequently, Euclidean geometry cannot be taken to presuppose the laws of motion. The direction of explanation goes the other way around. It is the principle of inertia that involves, or presupposes only very limited spatiotemporal features, namely that motion is uniformly (with regard to time) and rectilinearly (with regard to space).

As Garbe rightly notes, Lange did not go so far as to formulate the concept of relative distant simultaneity. Only the relativity of simultaneity makes possible the invariance of the velocity of light, or, to put it another way: the transformations between inertial frames that preserve the velocity of light will not preserve simultaneity. 

Garbe transfers Lange's idea of ``partial determination'' that still adheres to the notion of absolute simultaneity to the interpretation of Einstein's theory of relativity. According to Garbe, Einstein's theory of relativity is based on the empirical fact of the invariance of the velocity of light and on the convention that light signals define simultaneity. Both aspects together make Einstein's theory so attractive for ``partial determination.'' 

This holds for special relativity as well as for general relativity. Special relativity shares with Newtonian mechanics one crucial point: inertial coordinates are distinguished from all others (i.\,e. from non-inertial reference frames), and the laws of physics are said to hold only relative to inertial coordinate systems. It is often argued that general relativity puts away any worries over inertial reference frames insofar as it nullifies the privileged status of inertial reference frames. Inertial and uniformly accelerated motions are equivalent. However, the statement that all reference frames, rather than just inertial frames, are equivalent is misleading. Rather, the variable curvature of space-time makes the imposition of a global inertial frame impossible. Any space-time obeying the general theory of relativity and thus accounting for gravitation will be locally Minkowskian in the sense that any infinitesimal region of space-time has an inertial frame obeying the principles of special relativity.

General relativity teaches us the lesson that the space-time distribution of the energy-momentum tensor determines the local geometry encoded in the metric tensor, and the Ricci tensor is determined by the metric tensor and its space-time derivatives. Einstein's field equations can be written in the form:\\

\begin{equation}
R_{\mu\nu} \equiv R_{\mu\nu} - \frac{1}{2} Rg_{\mu\nu} = \frac{8 \pi G}{c^{4}} T_{\mu\nu}
\end{equation}

\noindent

On the left hand side is the Einstein tensor $G_{\mu\nu}$, a specific divergence-free combination of the Ricci curvature tensor $R_{\mu\nu}$ and the metric, where $R$ is the scalar curvature and $g_{\mu\nu}$ is the metric tensor, $G$ is Newton's gravitational constant, $c$ is the speed of light in vacuum. On the right hand side, $T_{\mu\nu}$ is the energy-momentum tensor. By means of this equation the distribution of matter and energy is combined with the curvature of space-time. 

The crucial question is how one should interpret this connection. According to Garbe, one should exercise caution in embracing the conclusion that the distribution of mass/energy causes space-time and that this curvature causes the gravitational field. Einstein's equations relate the stress energy tensor to the metric tensor. Curvature is then determined by the metric and its derivatives. But this does not imply a causal (and temporal) direction of determination. Rather, gravitational potentials were coordinated to the pseudo Riemannian space-time metric. Due to this coordination physics and geometry were connected with each other by conventional procedures. The law which connects the metric of the Riemannian space-time with the sources of the gravitational field (including the boundary/initial conditions) is given by the field equations. These equations have been approximately confirmed by astronomical observations concerning the paths of light rays in the gravitational field of the sun. That's what Garbe means when she speaks of ``partial determination.''

One lesson we can learn from Garbe's book is that conventions function as a priori concepts in a liberal and pragmatic sense. On the one hand, they are not immune to revision. On the other hand, we are not free to regard these concepts as definitions or empirical concepts; instead, their constitutive status is determined by the fact that they make certain kinds of empirical tests possible. 

It is not difficult to see that Garbe's approach comes very close to Michael Friedman's contemporary defense of the relativized a priori. Whereas Friedman's analysis of the role of conventional and constitutive principles in the theory of relativity initiated much discussion in the literature, Garbe's defense of ``partial determination'' has been ignored by the scientific community. This is entirely unjust due to her highly original re-discovery of Ludwig Lange.

Finally, just one last sceptical comment: Garbe combines her position with a skeptic attitude towards causal interpretations of Einstein's field equations. She argues that it is misleading to claim that gravity is caused by curved space-time. From my point of view, it is problematic to interpret the field equations of general relativity as claiming that the stress energy tensor causes the space-time metric, not because of the conventional infiltration, but because the relationship between the terms on the two sides of the field equations does not represent an \textit{asymmetric} determination we associate with causation, but an interaction between matter and the metric field. To put is more precisely, not only are space-time and matter in dynamical interaction, but the metric field itself is both a source of its own curvature and of a characteristic form of radiation, i.\,e. gravitational waves.

\flushright\copyright\, 2016\, Andrea Reichenberger, University of Paderborn

\end{document}